\documentclass[11pt,a4paper]{article}
\usepackage{jheppubnohead}
\usepackage[utf8]{inputenc}
\usepackage{epsfig}
\usepackage{tabularx}
 \usepackage{amsmath,amsfonts,amssymb,dsfont,mathrsfs,graphicx, lieart, booktabs, siunitx, caption, subcaption}
\usepackage[capitalise, english]{cleveref}

\def\beqn{\begin{eqnarray}}
\def\eeqn{\end{eqnarray}} 
\def\be{\begin{equation}}
\def\ee{\end{equation}}

\def\UPQ{\text{U}(1)_{\rm PQ}}
\DeclareMathOperator{\STr}{STr}

\def\lxone{\lambda_{x_{11}}}
\def\lxtwo{\lambda_{x_{22}}}
\def\lxthree{\lambda_{x_{12}}}
\def\lxfour{\lambda_{x_{21}}}

\def\lsone{\lambda_{s_{11}}}
\def\lstwo{\lambda_{s_{22}}}
\def\lsthree{\lambda_{s_{12}}}
\def\lsfour{\lambda_{s_{21}}}

\def\mc{\mathcal}

\newcommand{\rcharge}[1]{q_{#1}^R}

\newcommand{\dcharge}[1]{q_{#1}}

\newcommand{\Mod}[1]{\ (\text{mod}\ #1)}

\makeatletter
\newcommand{\pushright}[1]{\ifmeasuring@#1\else\omit\hfill$\displaystyle#1$\fi\ignorespaces}
\newcommand{\pushleft}[1]{\ifmeasuring@#1\else\omit$\displaystyle#1$\hfill\fi\ignorespaces}
\makeatother
\makeatletter
\newcommand{\specialcell}[1]{\ifmeasuring@#1\else\omit$\displaystyle#1$\ignorespaces\fi}
\makeatother

\makeatletter
\providecommand*{\diff}%
	{\@ifnextchar^{\DIfF}{\DIfF^{}}}
\def\DIfF^#1{%
	\mathop{\mathrm{\mathstrut d}}%
		\nolimits^{#1}\gobblespace}
\def\gobblespace{%
	\futurelet\diffarg\opspace}
\def\opspace{%
	\let\DiffSpace\!%
	\ifx\diffarg(%
		\let\DiffSpace\relax
	\else
		\ifx\diffarg[%
			\let\DiffSpace\relax
		\else
  			\ifx\diffarg\{%
				\let\DiffSpace\relax
			\fi\fi\fi\DiffSpace}

 \title{Dynamical generation of the Peccei-Quinn scale in gauge mediation}
 
 \author[a]{Guido~Festuccia,}
 \author[b]{Toby~Opferkuch,}
 \author[c]{Lorenzo~Ubaldi}

 \affiliation[a]{Niels Bohr International Academy and Discovery Center, Niels Bohr Institute, \\
  University of Copenhagen, Blegdamsvej 17, 2100 Copenhagen \O, Denmark}
  \affiliation[b]{Physikalisches Institut der Universit\"at Bonn, Bethe Center
for Theoretical Physics, \\ Nu{\ss}allee 12, 53115 Bonn, Germany}
\affiliation[c]{Raymond and Beverly Sackler School of Physics and Astronomy, \\
 Tel-Aviv University, Tel-Aviv 69978, Israel}

\emailAdd{guido@scipp.ucsc.edu}
\emailAdd{toby@th.physik.uni-bonn.de}
\emailAdd{ubaldi.physics@gmail.com}

\abstract{The Peccei-Quinn (PQ) mechanism provides an elegant solution to the strong CP problem. 
However astrophysical constraints on axions require the PQ breaking scale to be far higher than the electroweak scale. In supersymmetric models the PQ symmetry can be broken at an acceptable scale if the effective potential for the pseudo-modulus in the axion multiplet develops a minimum at large enough field values. In this work we classify systematically hadronic axion models in the context of gauge mediation and study their effective potentials at one loop. We find that some models generate a PQ scale comparable to the messenger scale. Our result may prove useful for constructing full realistic models of gauge mediation that address the strong CP problem. We also comment briefly on the cosmological aspects related to saxion and axino, and on the quality of the PQ symmetry. }

\begin{document}
\hfill {\tt BONN-TH-2015-06}

\maketitle
\section{Introduction}
The Peccei-Quinn (PQ) mechanism~\cite{Peccei:1977hh, Peccei:1977ur} is one of the most plausible solutions to the strong CP problem~\cite{Peccei:2006as}. It involves an anomalous $\UPQ$ symmetry that is spontaneously broken. The corresponding pseudo-goldstone boson is the axion~\cite{Wilczek:1977pj, Weinberg:1977ma}. Astrophysical constraints~\cite{Raffelt:2006cw} on axions imply that the PQ breaking scale has to be very high ($f_a > 10^9$ GeV). This introduces a large hierarchy between the PQ and the electroweak scale which calls for a theoretical explanation. One way to address the issue is in the context of Supersymmetry (SUSY), using the SUSY breaking effects to generate the PQ scale dynamically.

This possibility was first considered by two different groups~\cite{Asaka:1998ns,ArkaniHamed:1998kj}. They proposed a supersymmetric hadronic axion model composed of three sectors: a hidden sector where SUSY is broken, a PQ sector defined by the superpotential 
\be
W_{\rm PQ} = \lambda_p S \tilde \Phi_p  \Phi_p  \, ,
\ee
 and a messenger sector
 \be
 W_M = \lambda_M X  \tilde \Phi_M  \Phi_M \, .
 \ee
The chiral superfields $S$, $\Phi_p$ and $\tilde\Phi_p$ are charged under the $\UPQ$, with $S$ the gauge-singlet axion multiplet, and $\Phi_p,\;\tilde\Phi_p$ in the $\irrep{3}$ and $\irrepbar{3}$ of the color gauge group $\text{SU}(3)_C$. The superfield $X$ has a  SUSY breaking vacuum expectation value (VEV), $\langle X \rangle = M + \theta^2 F$, and $\Phi_M,\; \tilde\Phi_M$ are the messenger superfields of gauge mediation, also in the $\irrep{3}$ and $\irrepbar{3}$ of $\text{SU}(3)_C$.  At two loops gauge mediation generates SUSY breaking masses for the PQ squarks $\Phi_p$ and $\tilde{\Phi}_p$. In turn SUSY breaking effects are transmitted to the scalar $S$ field via one-loop diagrams of $\Phi_p$ and $\tilde{\Phi}_p$. This generates an effective potential $V_{\rm eff}(S)$, effectively at three loops, such that $|S| \frac{\partial V_{\rm eff}}{\partial \tilde S} < 0$. The effects of gravity mediation, parametrized as
 \be
 V_{\rm gr} \sim m_{3/2}^2 |S|^2 \, ,
 \ee
 with $m_{3/2}$ the gravitino mass, then stabilize the potential for $S$ at large field values. This results in a large PQ breaking scale, $f_a = \langle S \rangle$. 
 
 A similar setup was reexamined more recently in the context of gauge mediation. 
 Some of the present authors considered a model in which the PQ messengers $\Phi_p, \tilde\Phi_p$ and the regular messengers $\Phi_M, \tilde\Phi_M$ mixed~\cite{Carpenter:2009sw} and concluded that at one loop it did not stabilize the PQ scale away from the origin of the field $S$.  The same work~\cite{Carpenter:2009sw} considered other ways of generating the PQ scale dynamically, but they all involved models with additional gauge interactions that were somewhat complicated.
 In Ref.~\cite{Jeong:2011xu} the authors considered a slightly different model with messenger mixing
 and showed that it led to the stabilization of the PQ scale at two loops.\footnote{See also Refs.~\cite{Choi:2011rs,Nakayama:2012zc} for different mechanisms of stabilization of the PQ scale in gauge mediation.} 
  
 In this work we revisit the dynamical generation of the PQ scale in the context of gauge mediation and we ask if it is possible to achieve it in a simpler way.
To this end we consider models with an $R$-symmetry and a PQ symmetry, containing the fields $S$, $X$ and an arbitrary number of messengers. We classify these models systematically according to the charge assignments of the fields, and
point out possible virtues and disadvantages for each class.
We study in detail concrete examples in which the one-loop effective potential generates a VEV, $\langle S \rangle \sim M$, with $M$ the messenger scale. Taking a high enough messenger scale, $M > 10^9$ GeV, we then have an acceptable PQ breaking scale, $f_a \sim M$. This is significantly simpler than the scenarios we described above. First, the analysis at one loop is sufficient. Second, the PQ scale is stabilized solely within the context of gauge mediation, without the need of intervention of gravity mediation.
 
The paper is organized as follows. In section~\ref{sec:models} we introduce the models and categorize them into four distinct classes.  In \cref{sec:completemodel} we study a couple of examples and comment on the vacuum stability and on cosmological constraints.
In \cref{sec:quality} we discuss how one could address the issue of the axion quality in these models by invoking discrete symmetries.
In section~\ref{sec:discussion} we summarize our main results and discuss possible implications for phenomenological model building. 
We include two appendices. In the first we report many details on models with two sets of messengers. In the second we study the anomalies related to the discrete gauge symmetries, invoked to address the axion quality problem.


\section{Classification of models} \label{sec:models}
We consider models defined by the superpotential
\be \label{eq:superdef}
W = \mathcal{M}_{ij}(X,S) \tilde\Phi_i \Phi_j + W_R(X) \, ,
\ee
where
\be \label{eq:bigM}
\mathcal{M}_{ij}(X,S) = X \lambda_{ij} + m_{ij} + S \delta_{ij} \, , \qquad i,j = 1,\dots, N \, .
\ee
There are $N$ messenger fields, $\Phi_i,\tilde \Phi_i$, transforming under $\irrep{5}$ and $\irrepbar{5}$ of $\text{SU}(5)$ respectively, and two gauge-singlet fields, $X$ and $S$. The matrices $\lambda$, $m$ and $\delta$ are $N\times N$. We assume a global $R$-symmetry, $U(1)_R$, and a PQ symmetry, $\UPQ$, under which the fields are charged as shown in \cref{tab:U1charges}. The choice of keeping $X$ neutral under $\UPQ$ is convenient as it keeps the SUSY breaking and PQ breaking sectors separate. Note also that our models, as most models of gauge mediation, have an extra $U(1)_V$ global symmetry, under which the $\Phi$'s and the $\tilde \Phi$'s transform with opposite phases. The quantum number associated with this symmetry is the messengers' number.\footnote{If the messengers' number were conserved the lightest component of the messengers would be stable and could overclose the universe. We assume that $U(1)_V$ is broken in another sector of the theory. A coupling of one messenger to two matter fields in the superpotential of the visible sector, for instance, would suffice~\cite{Evans:2013kxa}.}

The term $W_R(X)$ in \cref{eq:superdef} stands for a hidden sector superpotential, necessary to generate a SUSY breaking VEV for $X$.
A minimal choice is $W_R(X)= F X$ but for the time being we leave $W_R(X)$ unspecified.  
Consistent with our choice that $X$ is uncharged under the PQ symmetry, we assume that all the fields that appear in $W_R(X)$ are singlets under $\UPQ$.

 Our notation, as well as the reasoning we outline in the rest of this section, are inspired by the models of (extra)ordinary gauge mediation~\cite{Cheung:2007es} where the superpotential has the form of~\cref{eq:superdef} with $S$ absent.
\begin{table}
\begin{center}
\begin{tabular}{l l l l l} \toprule
	{} & $X$ & $S$ & $\Phi_i$ & $\tilde\Phi_i$  \\\midrule
	$R$ & $2$ & $r_S$ & $r_i$ & $\tilde r_i$ \\
	${\rm PQ}$ & $0$ & $p_S$ & $p_i$ & $\tilde p_i$ \\
	\bottomrule	 
\end{tabular} 
\end{center}
\caption{$R$ and PQ charges of the fields in the model.} 
\label{tab:U1charges}
\end{table}
The superpotential in \cref{eq:superdef} must have $R$ charge 2 and PQ charge zero, resulting in the following selection rules
\begin{align}
& \lambda_{ij} \neq 0 \quad & {\rm only} \ {\rm if} \quad  &\tilde r_i + r_j = 0 \quad & {\rm and} & \quad \tilde p_i +  p_j = 0 \, , \label{eq:rulelambda} \\
& m_{ij} \neq 0 \quad & {\rm only} \ {\rm if} \quad  & \tilde r_i +  r_j = 2 \quad & {\rm and} & \quad  \tilde p_i +  p_j = 0 \, , \label{eq:rulem} \\
& \delta_{ij} \neq 0 \quad & {\rm only} \ {\rm if} \quad  & \tilde r_i +  r_j = 2-r_S \quad & {\rm and} & \quad \tilde p_i + p_j = -p_S \, . \label{eq:ruledelta}
\end{align}
As a consequence the determinant of the matrix $\mathcal{M}$ is a monomial in $X$ and $S$:
\be \label{eq:monomials}
\det \mathcal{M} = X^n S^q G(\lambda, m, \delta) \, ,
\ee
where $G(\lambda, m, \delta)$ is some function of the couplings and
\begin{align}
n & = \sum_{i=1}^N \left(1 - \frac{1}{2}( \tilde r_i +r_i) + \frac{r_S}{2} \frac{(\tilde p_i + p_i)}{p_S} \right) \, , \\
q & = -\frac{1}{p_S} \sum_{i=1}^N ( \tilde p_i + p_i ) \, .
\end{align}
Here $n$ and $q$ are integers satisfying $0 \leq n \leq N$ and $0 \leq q \leq N$. The proof of the identity in \cref{eq:monomials} can be done in analogy with that given in Ref.~\cite{Cheung:2007es}.

Our aim is to look for models that generate a large VEV for the field $S$, so that the PQ symmetry is spontaneously broken. Note that this implies that the PQ charge of $S$ must be strictly nonzero, $p_S \neq 0$. As $S$ is a flat direction of the classical potential we have to compute the one-loop effective potential and check if it stabilizes $S$ away from the origin. The identity of \cref{eq:monomials} leads to a classification scheme with four qualitatively distinct types of models.

\begin{itemize}
\item {\em Type A:} $\det m \neq 0$, $\det \lambda = \det \delta = 0$.

Here we can go to a basis where $m$ is diagonal, which implies $\tilde r_i +r_i = 2$ and $\tilde p_i +p_i = 0$. It follows that
\be
 n = q = 0 \, .
\ee
The messengers are stable around $X=0$ and $S=0$, but some can become tachyonic at large $X$ and $S$.
In these models the gaugino masses vanish to leading order in $F$. We will not study any models of this kind in the remainder of the paper.

\item {\em Type B:} $\det \lambda \neq 0$, $\det m = \det \delta = 0$.

Here we can go to a basis where $\lambda$ is diagonal, $ \tilde r_i +r_i = 0$ and $ \tilde p_i +p_i = 0$. Thus we have
 \be
 n=N \quad  {\rm and} \quad q=0 \, . 
 \ee
 At large $X$ all messengers have masses of order $\lambda X$. As $X$ approaches the origin, $\det m = 0$ implies that some messengers have $\mathcal{O}(m)$ masses, while others are light with masses going to zero as some power of $X$. Eventually the latter become tachyonic as $X$ gets closer to zero. There can however be local minima of the potential for finite $X$ where all the messengers are massive. We study a simple model of this form in \cref{sec:twomess}.

\item {\em Type C:} $\det \delta \neq 0$, $ \det \lambda =\det m = 0$.

Here we can go to a basis where $\delta$ is diagonal, $ \tilde r_i +r_i = 2 - r_S$ and $ \tilde p_i +p_i = -p_S$. It follows that
\be
n=0 \quad {\rm and} \quad q=N \, .
\ee
With a reasoning analogous to the one for Type B models, we deduce that in Type C models some messengers become tachyonic as $S$ goes to zero. As $n=0$, the gauginos are massless at leading order in the SUSY breaking parameter $F$. In the next section we study in detail a model belonging to this category for which $S$ is stabilized away from the origin.

\item {\em Type D:} $ \det \lambda =\det m = \det \delta= 0$.

In this category we have
\be
 0 < n < N \quad {\rm and} \quad 0 < q < N \, .
 \ee
  The messenger sector combines features of the previous categories and there are no tachyons in a region $X_{\rm min} < |X| < X_{\rm max}$, $S_{\rm min} < |S| < S_{\rm max}$. We will see that models of this kind can develop a minimum with $\langle X\rangle\neq 0$ and $\langle S\rangle\neq 0$ at one loop. 

\end{itemize}


\section{Concrete examples} \label{sec:completemodel}
\subsection{A model with two sets of messengers}
In this section we consider a model of Type C with two sets of messengers ($N=2$) that generates a VEV for the field $S$ of order the messenger scale $M$. Taking $M>10^9$ GeV we then have an acceptable PQ breaking scale. Many details of the analysis are found in \cref{sec:twomess} where we look in general at models with $N=2$ and find other options (of Type B and D) with less appealing features. 
For the sake of definiteness we specify $W_R(X)$, which we take to be of the simple form proposed in Ref.~\cite{Shih:2007av}.
The superpotential of the model is then given by:
\begin{align} 
W & = W^C_{\rm PQ} + W_R \, ;  \label{eq:fullmod} \\
W^C_{\rm PQ} &  = X \lambda_x \tilde\Phi_1 \Phi_1 + S  \lambda_s \left( \tilde \Phi_1 \Phi_2 + \tilde\Phi_2 \Phi_1   \right) \, ,  \label{eq:WC} \\
W_R & = X(\lambda \varphi_1 \varphi_{-1} + F)  + m_1 \varphi_{-1} \varphi_3 +\frac{1}{2} m_2 \varphi_1^2 \, . 
\end{align}
Here the $\varphi$ fields are gauge- and PQ-singlets, with the subscript denoting their $R$ charge. The PQ and $R$ symmetries forbid additional renormalizable terms in~\cref{eq:fullmod}. 
We have taken the two couplings of $S$ to the messengers to be equal. With this choice the model has a messenger parity
symmetry~\cite{Dimopoulos:1996ig} that has the virtue of forbidding dangerous D-terms~\cite{Dine:1981gu}, which would otherwise lead to tachyonic sfermions.  

At tree level there is a pseudo moduli space of vacua on which $\varphi_i=0,~\Phi_i=0,~\tilde \Phi_i=0$ with $X$ and $S$ arbitrary. As described in the previous section some of the messengers are tachionic for small $S$. Indeed for small $S$ the potential rolls down to a moduli space of supersymmetric vacua with $\varphi_i=0,\;S=0,\;X=0$ and on which the gauge invariant combinations $\tilde\Phi_i  \Phi_j$ are subject to the constraints:
\be
\tilde \Phi_1  \Phi_1=-{F\over \lambda_x}~,\qquad \tilde\Phi_1 \Phi_2+\tilde\Phi_2  \Phi_1=0~.
\ee
Moreover on the pseudo moduli space  some of the $\varphi_i$ become tachionic at large $X$. In this case the potential rolls down  along a runaway direction on which $ \Phi_i=\tilde \Phi_i=0$. This runaway is parametrized by $\varphi_3\rightarrow \infty$ and~\cite{Shih:2007av}
\be
X = \left({m_1^2 m_2 \varphi_3^2\over \lambda^2 F}\right)^{1\over 3},\qquad \varphi_1=\left({F m_1\varphi_3\over \lambda m_2}\right)^{1\over 3},\qquad \varphi_{-1}=\left({F^2 m_2\over \lambda^2 m_1 \varphi_3}\right)^{1\over 3} ~.
\ee

We are interested in establishing if the one loop effective potential on the pseudo moduli space has a local minimum in $X$ and $S$ such that no field is tachionic. The Coleman-Weinberg formula for the potential at one loop is:
\be \label{eq:CWfull}
V^{(1)} = \frac{1}{64 \pi^2} \STr \left[ \hat M^4 \left(\log \frac{\hat M^2}{\Lambda^2} -\frac{1}{2} \right) \right] \, ,
\ee
where $\STr$ stands for supertrace, $\hat M^2$ is shorthand for the squared mass matrices of the scalar and fermion components of the superfields and $\Lambda$ is the cutoff scale. We can compute the one-loop effective potential $V^{(1)}(X,S)$, as a function of both $X$ and $S$, at all orders in the SUSY breaking parameter $F$.
Given the form of $W$ in \cref{eq:fullmod} the squared mass matrices of scalars and fermions are block diagonal and the resulting one-loop potential can be written as the sum of two terms
\be \label{eq:V1factors}
V^{(1)}(X,S) = V^{(1)}_{\rm PQ}(X,S) + V^{(1)}_R(X) \, ,
\ee
where $V^{(1)}_R(X)$ was computed in Ref.~\cite{Shih:2007av}. For some range of the parameters one obtains a minimum at $\langle X \rangle =M$ for which all the $\varphi_i$'s are non tachyonic.
The $V^{(1)}_{\rm PQ}(X,S)$ contribution to the one loop potential will not destabilize this minimum provided that 
\be \label{eq:Xmincond}
\frac{\partial V^{(1)}_{\rm PQ}(X,S)}{\partial X} \ll \langle X \rangle \frac{\partial^2 V^{(1)}_R(X)}{\partial X^2} \, .
\ee
This condition can be satisfied by taking $\lambda_x$ sufficiently small, $\lambda_x < 0.1 \ \lambda$.
 
In \cref{sec:twomess} we present a detailed analysis of $V^{(1)}_{\rm PQ}(X,S)$.  In summary there is a local, PQ breaking minimum at 
\be \label{eq:modmin}
S_{\rm min} \simeq \frac{\lambda_x}{\lambda_s} \ e^{-3/2} M \, .
\ee
 For $S < S_{\rm tac}$, with $S_{\rm tac} = \frac{\sqrt{\lambda_x F}}{\lambda_s}$, some messengers become tachyonic and the system rolls down classically to the SUSY vacuum. For $S_{\rm min}$ to lie in the tachyon-free region we need to satisfy
\be
F < e^{-3} \lambda_x M^2 \, .
\ee

We want to make sure that our metastable vacuum is long lived, {\it i.e.} that the tunnelling rate from $S_{\rm min}$ to $S = 0$ is low\footnote{The tunneling rate to the runaway direction where $\varphi_3\rightarrow \infty$ can be easily made very small~\cite{Shih:2007av}.}. A detailed study of this rate is beyond the scope of this work. It is sufficient to note that the potential barrier height does not vary strongly with $\lambda_s$ while its width, $S_{\rm min} - S_{\rm tac}$, is proportional to $\lambda_s^{-1}$. Hence the tunneling rate can be made small by lowering $\lambda_s$. 

At the metastable minimum of $V^{(1)}(X,S)$, with $\langle S \rangle \sim \langle X \rangle = M$, we have the following spectrum:
\begin{enumerate}
\item The axion and the $R$-axion are massless.\footnote{The axion then acquires a small mass due to the $\UPQ$ anomaly, while the $R$-axion can acquire a mass from explicit $R$-symmetry breaking in supergravity~\cite{Bagger:1994hh}. }
\item The saxion has a mass
\be \label{eq:saxmass}
m_s \sim \sqrt{\frac{\lambda_s^2}{16 \pi^2}} \frac{F}{M} \, .
\ee 
In terms of loop counting this is larger than that of the MSSM particles in gauge mediation. However vacuum stability requires small $\lambda_s$, and as a result the saxion is likely lighter than most of the MSSM particles.
\item The axino, the fermionic component of the superfield $S$, has a mass
\be \label{eq:axinomass}
m_{\tilde a} \sim \frac{\lambda_s^2}{16 \pi^2} \frac{F}{M} \, ,
\ee  
so it is lighter than the saxion.
\item The $R$-saxion has a mass
\be \label{eq:Rsaxmass}
m_{Rs} \sim \sqrt{\frac{\lambda^2}{16 \pi^2}} \frac{F}{M} \, ,
\ee 
and is typically heavier than the saxion, as the coupling $\lambda$ that appears in $W_R$ does not need to be as small as $\lambda_s$.
\item The gravitino is light, as usual in gauge mediation
\be
m_{3/2} \sim \frac{F}{M_{\rm Pl}} \, ,
\ee
where $M_{\rm Pl}$ is the reduced Planck mass.
\end{enumerate}

The saxion and $R$-saxion are the two pseudomoduli in the model and we need to make sure that they do not pose cosmological issues~\cite{Banks:2002sd}. Typically such issues are more severe the lighter the pseudomodulus, so we concentrate on the saxion here, as $m_s < m_{Rs}$. The main decay modes of the saxion are (i) into two axions, (ii) into an axino and a gravitino, and (iii) into two gravitinos.\footnote{The saxion can also decay into pairs of MSSM particles or pairs of gauge bosons, but these are further suppressed and subdominant compared to (i), (ii), and (iii) in the text.}  The decays (ii) and (iii) can be understood from the effective operator
\be \label{eq:highK}
\frac{1}{M^2} \int d^4 \theta \ (X^\dagger X) (S^\dagger S) \, .
\ee 
The decay rate in each case is given approximately by
\be \label{eq:decay}
\Gamma_s \sim \frac{1}{16 \pi} \frac{m_s^3}{M^2} \simeq 10^{-25} \ {\rm GeV} \ \left( \frac{m_s}{1 \ {\rm GeV}} \right)^3 \left( \frac{10^{12} \ {\rm GeV}}{M} \right)^2 \, .
\ee  
The saxion starts oscillating about its minimum when the Hubble rate, $H \sim T^2/M_{\rm Pl}$, is comparable to its mass, $H \sim m_s$. At that time it has an energy density of order $m_s^2 M^2$, and constitutes a fraction $M^2/M_{\rm Pl}^2$ of the total energy density. From then on it behaves like matter, so its fraction of the energy density grows with the scale factor. It decays when $H \sim \Gamma_s$, that it is at a temperature
\be \label{eq:Tdecay}
T_s^{\rm dec} \sim \frac{m_s^{3/2} M_{\rm Pl}^{1/2}}{10 \ M} \, .
\ee
Requiring that it decays safely before Big Bang Nucleosynthesis (BBN), $T_s^{\rm dec} > 0.1$ GeV, amounts to a lower bound on $\lambda_s F$:
\be \label{eq:BBNconst}
\lambda_s F > 10^{15} \ {\rm GeV}^2 \left( \frac{M}{10^{12} \ {\rm GeV}} \right)^{5/3} \, ,
\ee
where we used \cref{eq:saxmass}. As long as $M<10^{12}$ GeV, the bound of \cref{eq:BBNconst} also guarantees that the saxion decays before it comes to dominate the energy density. This is helpful in relieving possible constraints from extra radiation at BBN~\cite{Kawasaki:2007mk}, as the decay products are relativistic.

The axino in this model is much heavier than the gravitino and its main decay mode is into a gravitino and an axion, which are both dark matter candidates. Avoiding over-closure of the universe with too much gravitino dark matter then results in a bound on the reheating temperature, which should be lower than about $10^5$ GeV~\cite{Cheung:2011mg}.

\subsection{Models with more messengers}
The model considered in \cref{eq:fullmod} is of Type C and has the deficiency that gaugino masses are suppressed, as they are not generated at one loop.
One way to fix this is to introduce more messengers. For example we can replace $W_R$ in \cref{eq:fullmod} with
\be
\label{eq:newbreak}
W_R=X \lambda (\tilde\Phi_3 \Phi_3+\tilde\Phi_4 \Phi_4 )+ X F + m \tilde \Phi_3 \Phi_4 \, .
\ee
Under the $R$-symmetry the new messengers have charges $\tilde r_4=r_3=1$ and $\tilde r_3=r_4=-1$.
The model specified by this $W_R$ plus $W^C_{\rm PQ}$ [which we take of the same form as in \cref{eq:WC}] is of Type D. 
 The superpotential (\ref{eq:newbreak}), which was considered in Ref.~\cite{Cheung:2007es}, breaks both SUSY and the $R$-symmetry spontaneously for a large range of parameters.  This continues to be true when added to the PQ sector as long as the couplings in $W^C_{\rm PQ}$ are not too large.\footnote{All possible couplings between the sets of messengers appearing in $W_{\rm PQ}^C$ and the messengers in $W_R$ need also be small.}
 This construction is rather ad hoc but serves as an example of a messenger sector that breaks SUSY, the $R$-symmetry and the PQ symmetry at one loop.

\section{The axion quality} \label{sec:quality}

The PQ mechanism provides a solution to the strong CP problem as long as the $\UPQ$ is violated almost exclusively by the QCD anomaly. One must ensure that higher-dimension operators which violate the symmetry explicitly~\cite{Kamionkowski:1992mf, Holman:1992us} are suppressed. In our framework such operators can appear in the superpotential, taking the form
\begin{align} \label{eq:superhigh}
\delta W \supset X M_{P}^2 \left(\frac{X}{M_{P}}\right)^a \left(\frac{S}{M_{P}} \right)^b \, , \qquad \text{where} \quad a \in \mathbb{N}^0\, , \, b \in \mathbb{N}^+\, ,
\end{align}
and $M_P$ is the Planck mass. These terms result in a contribution to the scalar potential that has to be small:
\begin{align} \label{eq:CPconstraint}
 |F| M_{P}^2 \left( \frac{f_a}{M_{P}}\right)^{a+b} < \SI{E-14}{\GeV^4}\, .
\end{align}
Here we have taken $\langle X \rangle \simeq \langle S \rangle = f_a \simeq M$. The above inequality must be satisfied if the axion is to provide the solution to the strong CP problem \cite{Carpenter:2009zs}. If $f_a \simeq \SI{E10}{\GeV}$ and $|F| \simeq \SI{E15}{\GeV^2}$, then this corresponds to the requirement that operators with $a+b < 8$ must be forbidden in the superpotential. 

Additionally one should worry about non-renomalizable operators in the K\"ahler potential. These are of the form
\begin{align} \label{eq:Khigh}
\delta K \supset \int \diff^2 \theta \diff^2 \bar{\theta} X X^\dag \left(\frac{X}{M_{P}}\right)^a \left(\frac{S}{M_{P}} \right)^b\, , \qquad \text{where} \quad a \in \mathbb{N}^0\, , \, b \in \mathbb{N}^+\, ,
\end{align}
which leads to the following contributions to the scalar potential
\begin{align}
F F^\dag  \left( \frac{f_a}{M_{P}}\right)^{a+b} < \SI{E-14}{\GeV^4}\, .
\end{align}
Taking once again $f_a \simeq \SI{E10}{\GeV}$ and $|F| \simeq \SI{E15}{\GeV^2}$, we must forbid operators with $a+b < 5$. This is a less stringent constraint on $a$ and $b$ compared to contributions from the non-renormalizable superpotential operators of \cref{eq:superhigh}.

One possibility to address this problem is to impose discrete gauge symmetries~\cite{Banks:1989ag,Krauss:1988zc,Ibanez:1991hv}. We consider imposing a discrete PQ symmetry, $\mathbb{Z}_N^{\rm PQ}$, and a discrete $R$-symmetry, $\mathbb{Z}_M^R$, such that the global $\text{U}(1)_R$ and $\UPQ$ described in the previous sections arise as accidental symmetries of the Lagrangian. The dangerous operators of \cref{eq:superhigh} and \cref{eq:Khigh} then have to respect $\mathbb{Z}_N^{\rm PQ} \times \mathbb{Z}_M^R$ and we have to find how large $N$ and
$M$ need to be in order to satisfy \cref{eq:CPconstraint}. The constraints arising from the requirement that these discrete symmetries are anomaly free are discussed in \cref{sec:anom}. The analysis is model dependent as it involves details of the visible sector (or other sectors) of the theory. As an example, for the model defined by \cref{eq:fullmod}, some of the possibilities are:
\begin{enumerate}
\item We can neglect anomaly constraints in our hidden sector model [\cref{eq:fullmod}] under the assumption that extra matter charged under $\text{SU}(5)$, $\mathbb{Z}_N^{\rm PQ}$ and $\mathbb{Z}_M^R$ is introduced to cancel anomalies~\cite{Harigaya:2013vja}. The minimal suitable discrete symmetry is then $\mathbb{Z}_3^{\rm PQ}\times \mathbb{Z}_5^R$, with the discrete $R$ charge of $S$ chosen as $\rcharge{S}=1$.
\item If there is no additional matter charged under $\mathbb{Z}_N^{\rm PQ}$ and $\text{SU}(5)$ we can still neglect anomaly constraints on the $\mathbb{Z}_M^R$ symmetry by assuming that the $R$ charges in the visible sector are chosen to cancel any resulting anomaly. The minimal suitable discrete symmetry is then $\mathbb{Z}_2^{\rm PQ}\times \mathbb{Z}_9^R$, with $\rcharge{S}=2$.
\item We can assume that the visible sector is anomaly free by itself. The minimal compatible discrete symmetry is then $\mathbb{Z}_2^{\rm PQ}\times \mathbb{Z}_{11}^R$, with $\rcharge{S}=5$.
\end{enumerate}

\section{Summary and discussion} \label{sec:discussion}

The feasibility of the axion solution to the strong CP problem requires a PQ breaking scale, $f_a$, much above the electroweak scale. It would be desirable to explain this hierarchy with a mechanism that generates $f_a$ dynamically. The main result of this paper is the construction of simple models that tie $f_a$ to the messenger scale $M$ of gauge mediation. This is achieved with a one-loop analysis where it is sufficient to compute the Coleman-Weinberg potential, induced by SUSY breaking effects, for the scalar component of the axion superfield $S$.
We find some examples in which such a potential leads to $\langle S \rangle \sim M$. This implies that the messenger scale, $M \sim f_a$, has to be larger than $10^9$ GeV, which puts it on the higher end of the range typically considered in gauge mediation.

 Our models possess an $R$-symmetry and a PQ symmetry, and can be classified in a way similar to Ref.~\cite{Cheung:2007es}. We find four distinct classes but study in detail only a few examples (some of which are in \cref{sec:twomess}). There is a lot left to explore. For instance, we have not studied any model of Type A. They suffer from the issue of suppressed gaugino masses, but perhaps one can address that problem in another sector of the theory. Models of Type D do not have such an issue and are possibly the most interesting to further investigate. We have mentioned only one example of a successful Type D model with four sets of messengers in \cref{sec:completemodel}. One can be more clever and find additional examples where the same sector spontaneously breaks the $R$-symmetry  (thus breaking SUSY) and the PQ symmetry. This would specify the entire ``hidden'' sector, which could then be connected to the visible sector in the gauge mediation framework. The result would be a full calculable model for which one could study the phenomenological implications. 
 
We have also addressed the problem of the axion quality in \cref{sec:quality}. We do not provide any new insight into this issue, and show that relatively large discrete symmetries, thus somewhat unattractive, are needed to ensure the high quality of the PQ symmetry.

\acknowledgments

We thank Michael Dine for very helpful conversations and insightful comments. 
G.F. is supported by a Mobilex grant from the Danish Council for Independent Research and FP7 Marie Curie Actions COFUND (grant id: DFF 1325-00061). L.U. is supported by the I-CORE Program of the Planning Budgeting Committee and the Israel Science Foundation (grant NO 1937/12).

\appendix

\section{Classification of models with two sets of messengers} \label{sec:twomess}
The simplest subcategory of the models we introduced in \cref{sec:models} consists of models with only two sets of messengers. As the goal is to generate a VEV for $S$ via the SUSY breaking effects encoded by $\langle X \rangle$, we need at least one nonzero entry for $\lambda_{ij}$ and one nonzero entry for $\delta_{ij}$. With $N=2$ this immediately implies $m = 0$, which means that models in this subcategory can only be of Type B, C or D.
In this Appendix we show that with this field content there is one model for each Type (except A) which is representative of any model one could write.

We assume a canonical K\"ahler potential for all the fields, and we write the superpotential as
\be \label{eq:super}
W = W_R(X) + W_{\rm PQ} \, .
\ee
We start from the most general $W_{\rm PQ}$ consistent with the $\text{SU}(5)$ and $R$ symmetries:
\begin{align}
W_{\rm PQ} & = X \left( \lxone \tilde\Phi_1 \Phi_1 + \lxtwo \tilde\Phi_2 \Phi_2 + \lxthree \tilde\Phi_1 \Phi_2 + \lxfour \tilde\Phi_2 \Phi_1 \right)  \label{Xterms} \\
                     & + S \left( \lsone \tilde\Phi_1 \Phi_1 + \lstwo \tilde\Phi_2 \Phi_2 + \lsthree \tilde\Phi_1 \Phi_2 + \lsfour \tilde\Phi_2 \Phi_1 \right) \, .  \label{Sterms}
\end{align}
We do not write the term $XF$ here, also consistent with the symmetries, but we assume it is included in $W_R(X)$. 
We take all the couplings $\lambda$'s to be real. We impose a global $\UPQ$ under which $S$ and the messengers are charged but $X$ is neutral:
\be \label{PQass}
X \to X \, , \quad S \to e^{i \alpha p_S} S \, , \quad \Phi_i \to e^{i \alpha p_i} \Phi_i \, , \quad \tilde\Phi_i \to e^{i \alpha \tilde p_i} \tilde\Phi_i \, .
\ee
We then have the following equations for the PQ charges:
\begin{align}
\tilde p_1 + p_1 &= 0  &  \tilde p_1 + p_1 &= -p_S \label{x1}\\
\tilde p_2 + p_2 &= 0   & \tilde p_2 + p_2 &= -p_S    \label{x2}\\
\tilde p_1 + p_2 &= 0   & \tilde p_1 + p_2 &= -p_S \label{x3}\\
\tilde p_2 + p_1 &= 0   & \tilde p_2 + p_1 &= -p_S   \label{x4}  \, .
\end{align}
Clearly the two equations on each line are mutually exclusive. If we satisfy three equations of either column, the remaining one of the same column is also automatically satisfied, which would allow only operators with either $X$ or $S$. We need at least one operator with $X$ and one with $S$ in order to generate the PQ scale through the SUSY breaking effects. Thus we can satisfy at most two equations of either column.

Suppose we keep the first two equations on the right column. Then it is easy to check that we can only satisfy either the last or the second to last equation on the left column. If we keep the second to last we have
\be
 X \lxthree \tilde\Phi_1 \Phi_2 + S  \left(  \lsone \tilde\Phi_1 \Phi_1 + \lstwo \tilde\Phi_2 \Phi_2   \right) \, .
\ee
Here we can relabel the unbarred fields $\Phi_1 \leftrightarrow \Phi_2$, so that the above is equivalent to
\be \label{eq:WPQ1}
W^C_{\rm PQ} = X \lxone \tilde\Phi_1 \Phi_1 + S  \left(  \lsthree \tilde\Phi_1 \Phi_2 + \lsfour \tilde\Phi_2 \Phi_1   \right) \, .
\ee
Keeping the equations on the right of \eqref{x3} and \eqref{x4} and the one on the left of \eqref{x1} is equivalent to keeping \eqref{x1} and \eqref{x2} on the right and \eqref{x3} on the left as we did above. Had we chosen instead \eqref{x1} and \eqref{x2} on the right and \eqref{x4} on the left, we could have relabelled $\tilde\Phi_1 \leftrightarrow \tilde\Phi_2$ and would have ended up again with the superpotential in \cref{eq:WPQ1}. This defines $N=2$ models of Type C, with $X$ coupled diagonally to only one set of messengers and $S$ coupled off-diagonally to both sets of messengers.

With a procedure analogous to the one just outlined, we find $N=2$ models of Type B:
\be \label{eq:WPQ2}
W^{B}_{\rm PQ} =   X  \left(  \lxone \tilde\Phi_1 \Phi_1 + \lxtwo \tilde\Phi_2 \Phi_2   \right) + S \lsthree \tilde\Phi_1 \Phi_2 \, .
\ee

Next, we can keep the left of \eqref{x1} and \eqref{x3} and the right of \eqref{x2} and \eqref{x4}, which leads to
\be 
X  \left(  \lxone \tilde\Phi_1 \Phi_1 + \lxthree \tilde\Phi_1 \Phi_2   \right) + S \left(  \lstwo \tilde\Phi_2 \Phi_2 + \lsfour \tilde\Phi_2 \Phi_1   \right) \, .
\ee
Here one of the operators is redundant. We have $p_1 = p_2$, so we can use the field redefinition
to get rid of the second operator involving $X$. After relabelling fields and couplings we have
\be \label{eq:WPQ3}
W^{D}_{\rm PQ} =   X  \lxone \tilde\Phi_1 \Phi_1  + S \left(  \lstwo \tilde\Phi_2 \Phi_2 + \lsfour \tilde\Phi_2 \Phi_1   \right) \, .
\ee
We could have chosen a different field basis. For example we could have used the field redefinition to get rid of one of the operators with $S$ instead. Of course the resulting physics would be equivalent. 

Keeping the left of \eqref{x1} and \eqref{x4} and the right of \eqref{x2} and \eqref{x3} we have
\be 
  X  \left(  \lxone \tilde\Phi_1 \Phi_1 + \lxfour \tilde\Phi_2 \Phi_1   \right) + S \left(  \lstwo \tilde\Phi_2 \Phi_2 + \lsthree \tilde\Phi_1 \Phi_2   \right) \, .
\ee
Here $\tilde p_1 = \tilde p_2$ and we can redefine the fields $\tilde\Phi_1$ and $\tilde\Phi_2$ to get rid of one of the operators. Again we get rid of the second one and we write
\be \label{eq:WPQ4}
X   \lxone \tilde\Phi_1 \Phi_1  + S \left(  \lstwo \tilde\Phi_2 \Phi_2 + \lsthree \tilde\Phi_1 \Phi_2   \right) \, .
\ee
This is equivalent to eq.~\eqref{eq:WPQ3} under the exchange $\Phi_i \leftrightarrow \tilde\Phi_i$.

To summarize, we have seen that different choices of the PQ charges lead to:
\beqn
W^{B}_{\rm PQ} &=&   X  \left(  \lxone \tilde\Phi_1 \Phi_1 + \lxtwo \tilde\Phi_2 \Phi_2   \right) + S \lsthree \tilde\Phi_1 \Phi_2 \\
W^C_{\rm PQ} &=& X \lxone \tilde\Phi_1 \Phi_1 + S  \left(  \lsthree \tilde\Phi_1 \Phi_2 + \lsfour \tilde\Phi_2 \Phi_1   \right) \\
W^{D}_{\rm PQ} &=&   X  \lxone \tilde\Phi_1 \Phi_1  + S \left(  \lstwo \tilde\Phi_2 \Phi_2 + \lsfour \tilde\Phi_2 \Phi_1   \right)   \, .
\eeqn

The scalar potential obtained from $W_{\rm PQ}$ has a pseudo-moduli space given by
\be
\Phi_i = \tilde\Phi_i = 0 \, , \qquad X \ {\rm and} \ S \ \ {\rm arbitrary} \, .
\ee
The flat direction along $X$ is lifted by the one-loop dynamics of $W_R(X)$. This in turn leads to lifting the flat direction along $S$. We can check if for any of our three models in this Appendix the one-loop potential in the $S$-field direction has a minimum which is not at the origin. That would lead to the spontaneous breaking of $\UPQ$. Also we would like the $S$-field VEV to be large in order to generate an acceptable PQ scale.
 
 \subsection{One-loop potentials at order $F^2$}\label{sec:CW}
Under the assumption that $F \ll M^2$, instead of using \cref{eq:CWfull} we can study the one-loop potential up to order $\mathcal{O}(F^2)$ with a simpler formula~\cite{Grisaru:1996ve}
\be \label{eq:Kahler}
V^{(1)}_{F^2} = - K_{\rm eff} = \frac{1}{16 \pi^2} \int d^4\theta \ {\rm Tr} \left( \mathcal{M}^\dagger \mathcal{M} \log \frac{\mathcal{M}^\dagger \mathcal{M}}{e \Lambda^2} \right) \, .
\ee
Here the mass matrix $\mathcal{M}$ is defined by
\be
W_{\rm PQ} = (\tilde\Phi_1 \  \tilde\Phi_2) \mathcal{M}
\begin{pmatrix}
\Phi_1 \\
\Phi_2
\end{pmatrix} \, .
\ee
$\mathcal{M} = \mathcal{M}(X,S)$ is a function of the fields $X$ and $S$. The computation of the potential of \cref{eq:Kahler} proceeds as follows:
\begin{enumerate}
\item compute the eigenvalues of $\mathcal{M}^\dagger \mathcal{M}$ and take the trace in \cref{eq:Kahler};
\item substitute $X$ with its VEV $\langle X \rangle = M + \theta^2 F$ and expand the integrand for small $F$ ($F\ll M^2$);
\item perform the $d^4\theta$ integration.
\end{enumerate} 
This method of calculating the one-loop potential allows us to examine each one of our models without specifying the explicit form of $W_R$. 

For the rest of this section $X$ and $S$ denote the scalar components of the corresponding superfields.
Once we determine $V^{(1)}_{F^2}(S)$ we want to check whether it produces a large $\langle S \rangle$. The expressions we obtain from eq.~\eqref{eq:Kahler} with arbitrary couplings $\lambda$'s are complicated enough not to allow us to calculate analytically the extrema of the potential. Therefore we proceed in two steps:
\begin{enumerate}
\item we expand $V^{(1)}_{F^2}(S)$ for $|S| \ll M$ and we check if we have a local maximum or minimum at $S = 0$;
\item we expand $V^{(1)}_{F^2}(S)$ for $|S| \gg M$. In all the models we find that the potential increases at large values of $S$.
\end{enumerate} 
If there is a maximum at $S = 0$ we conclude there must be a minimum at $S \neq 0$. If there is a minimum at $S =0$ it could be local, and in principle one could have a global minimum at $S \neq 0$, although this never turns out to be the case for the models we study.
\subsubsection{Type B Model}
\cref{eq:WPQ2} yields a mass matrix of the form
\be
\mathcal{M}^{B} = 
\begin{pmatrix}
\lxone X & \lsthree S \\
0 & \lxtwo X
\end{pmatrix}\, .
\ee
This model has been considered in Ref.~\cite{Carpenter:2009sw}.
The one-loop potential at order $F^2$ is
\begin{align}
	V^{(1)}_{F^2} &= \frac{F^2}{32\pi^2}\Bigg[ \frac{2C}{M^2B}+\left(\lxone^2+\lxtwo^2-\frac{D}{M^6B^{3/2}}\right)\ln\left(\frac{M^2}{2\Lambda^2}\left(A-B^{1/2}\right)\right)\notag\\
	&+\left(\lxone^2+\lxtwo^2+\frac{D}{M^6B^{3/2}}\right)\ln\left(\frac{M^2}{2\Lambda^2}\left(A+B^{1/2}\right)\right)\Bigg],
	\label{eq:full2}
\end{align}
where
\begin{align}
	A&=\lxone^2+\lxtwo^2+\frac{\lsthree^2 |S|^2}{M^2},\\
	B&=A^2-4\lxone^2\lxtwo^2,\\
	C&=M^2(\lxone^2-\lxtwo^2)^2(\lxone^2+\lxtwo^2)+\lsthree^2|S|^2(\lxone^4+6\lxone^2\lxtwo^2+\lxtwo^4),\\
	D&=(\lxone^2-\lxtwo^2)^4M^6+\lsthree^6|S|^6(\lxone^2+\lxtwo^2)+\lsthree^4|S|^4M^2(3\lxone^4-2\lxone^2\lxtwo^2+3\lxtwo^4)\notag\\
	&+3\lsthree^2|S|^2M^4(\lxone^2+\lxtwo^2)(\lxone^2-\lxtwo^2)^2.
\end{align}
For small $|S|$
\begin{align}
V^{(1)}_{F^2}(|S| \ll M)&= \frac{F^2}{16\pi^2}\left[ \lxone^2+\lxtwo^2+ \lxone^2 \ln\left(\frac{\lxone^2 M^2}{\Lambda^2}\right)+\lxtwo^2 \ln\left(\frac{\lxtwo^2 M^2}{\Lambda^2}\right)\right] \notag\\
&\:+\frac{\lsthree^4 F^2}{32\pi^2 } \left[ \frac{ \lxone^4-\lxtwo^4+2\lxone^2\lxtwo^2 \ln\left(\frac{\lxtwo^2}{\lxone^2}\right)}{\left( \lxone^2-\lxtwo^2\right)^3} \right] \frac{|S|^4}{M^4}+\mc{O}\left(\frac{|S|^6}{M^6}\right)\, ,
\label{eq:M2SmallSExp}
\end{align}
the first three derivatives are zero at $|S| = 0$ and
\begin{align}\label{eq:M24thderV}
\frac{\partial^4 V^{(1)}_{F^2}(|S| \ll M)}{\partial S^2 \partial S^{*2}}&=\frac{\lsthree^4 F^2}{8\pi^2 \lxone^2 M^4} c(y)  \,.
\end{align}
Here we have defined $y \equiv \lxtwo / \lxone$ and
\begin{align}
	c(y)&=\frac{y^4-2y^2 \ln y^2-1}{\left(y^2-1\right)^3}\,.
\end{align}
The function $c(y)$ is positive for any $y$, thus we have a local minimum at the origin.

\begin{figure}[!t]
\centering
  \includegraphics{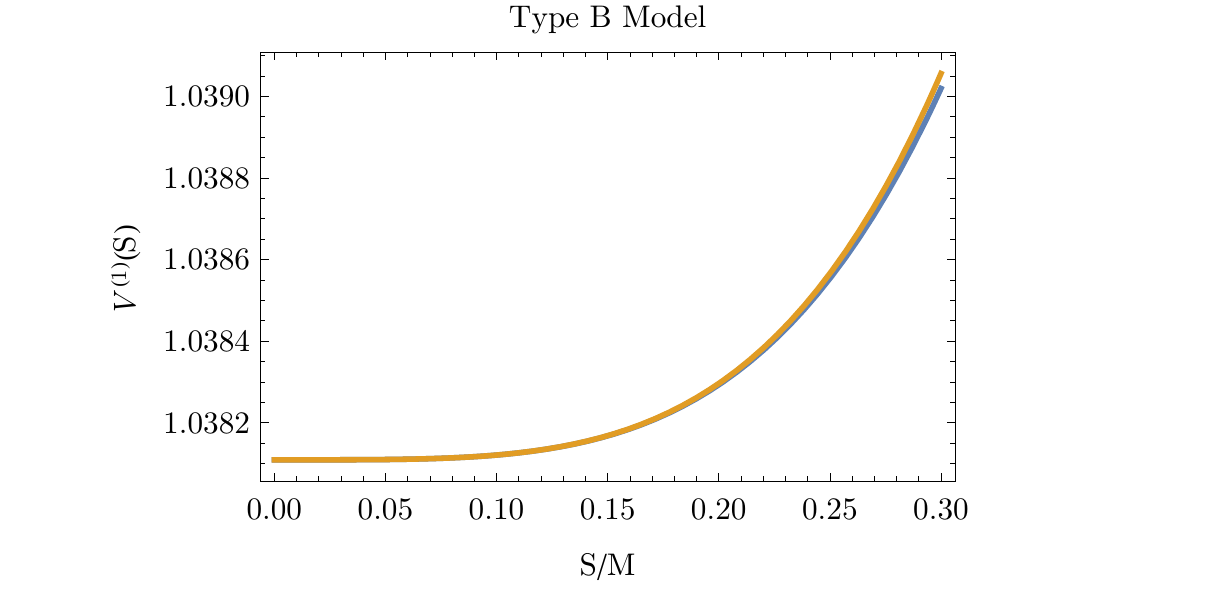}
\caption{A plot of the full K\"ahler effective potential (blue, \cref{eq:full2}) and the expansion of this potential around the origin (brown, \cref{eq:M2SmallSExp}) for the Type B model. We set $\lxone=\lsthree=1$, $\lxtwo=0.9$, $\Lambda=M$, and $F = 10^{-3} M^2$.}
\label{fig:M2SerVsFull}
\end{figure}

At large $|S|$ the potential grows logarithmically
\begin{align}
	V^{(1)}_{F^2}\left(|S| \gg M \right)&=\frac{F^2}{16\pi^2}\left(\lxone^2+\lxtwo^2\right)\ln\left(\frac{|S|^2 \lsthree^2}{\Lambda^2}\right)+\mc{O}\left(\frac{M^2}{|S|^2}\right)\,.
\end{align}
We can check graphically in \cref{fig:M2SerVsFull} that the minimum at the origin is the global one.
Thus this model does not seem to generate a VEV for the field $S$, as already stated in Ref.~\cite{Carpenter:2009sw}. Such a conclusion, however, is based on the analysis at order $F^2$. 

Let us see what happens when we consider higher orders in $F$. To perform the analysis, as we did in \cref{sec:completemodel}, we first have to specify $W_R(X)$. Again, we use the one proposed in Ref.~\cite{Shih:2007av}. Then we can use \cref{eq:CWfull} to compute the effective potential.
For the sake of simplicity we take all the couplings to be equal $\lxone=\lxtwo=\lsthree=\lsfour=1$.

\begin{figure}
\centering
\begin{minipage}{0.5\textwidth}
\centering
 \includegraphics{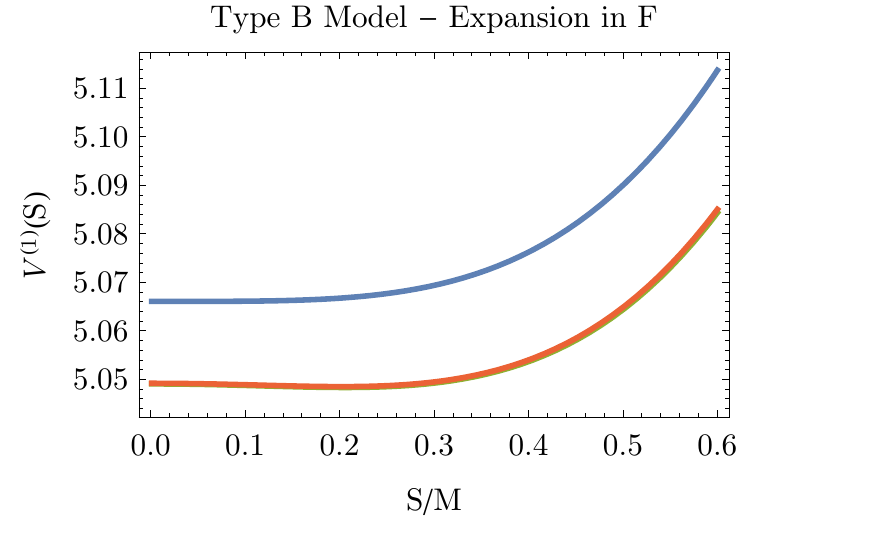}
\end{minipage}\hfill
\begin{minipage}{0.5\textwidth}
\centering
  \includegraphics{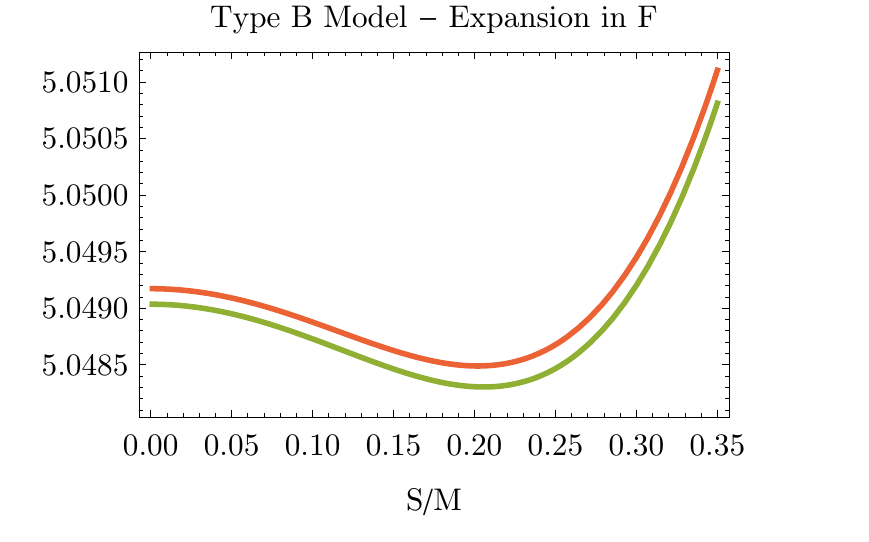}
\end{minipage}
\caption{The effective potential for model II. The green curve corresponds to the full effective potential, {\it i.e.} all orders in $F$, the blue is up to order $F^2$, the red is up to order $F^4$. On the left-hand plot the green curve lies almost directly behind the red curve. On the right-hand side we plot the same potential over a smaller domain: the minimum and the difference between the full potential and the $F^4$ expansion become apparent. Here a relatively large value of $F$ is chosen, namely $F=0.2 M^2$, and we set $\Lambda=M$.}
\label{fig:HigherF2}
\end{figure}

Shown in \cref{fig:HigherF2} is the full one-loop effective potential and its expansion to $\mathcal{O}(F^4)$. We see that there is indeed a minimum not located at the origin. This occurs only when including terms beyond $F^2$, and was missed in the analysis of Ref.~\cite{Carpenter:2009sw}. To find the location of the minimum, one can series expand the full one-loop potential both in $S$ and in $F$ yielding the following expression
\begin{align}\label{eq:MAFOLVsmall}
V^{(1)}(S\ll M) &= \frac{F^2}{8\pi^2}\Bigg[1+\frac{|S|^4}{12M^4}+\ln\left(\frac{M^2}{\Lambda^2}\right)\notag\\
&\quad\quad+\frac{F^2}{12M^4}\left(\frac{|S|^4}{10M^4}-\frac{2|S|^2}{ M^2}-1\right)\Bigg]+\mc{O}\left(\frac{|S|^6}{M^6}\right).
\end{align} 
Setting the first derivative with respect to $S$ to zero we find
\begin{align}
\langle |S| \rangle =  |S_{\text{min}}| &= \pm \frac{F}{M \sqrt{1+\frac{F^2}{10M^4}}} \simeq \pm\frac{F}{M}\,.
\end{align}
To obtain an acceptably high PQ scale in this model we need $F/M > 10^9$ GeV, which also leads to a very heavy supersymmetric particle spectrum.

\subsubsection{Type C Model}
This is the model we studied in detail in \cref{sec:completemodel}.
From \cref{eq:WPQ1} the mass matrix is
\be
\mathcal{M}^C = 
\begin{pmatrix}
\lxone X & \lsthree S \\
\lsfour S & 0
\end{pmatrix}\, .
\ee
The one-loop potential at order $F^2$ is
\begin{align}
	V^{(1)}_{F^2} &=\frac{\lxone^2 F^2}{32\pi^2}\Bigg[ \frac{2\lxone^2 A}{B}+\frac{\left(B^{1/2}-A\right)\left(B-\lxone^2 A- \lxone^2 B^{1/2}\right)}{B^{3/2}}\ln\left(\frac{M^2}{2\Lambda^2}\left(A-B^{1/2}\right)\right)\notag\\
	&+\frac{\left(B^{1/2}+A\right)\left(B-\lxone^2 A+\lxone^2 B^{1/2}\right)}{B^{3/2}}\ln\left(\frac{M^2}{2\Lambda^2}\left(A+B^{1/2}\right)\right)\Bigg]\,, \label{eq:full1}
\end{align}
where
\begin{align}
	A&= \frac{|S|^2(\lsthree^2+\lsfour^2)}{M^2}+\lxone^2,\\
	B&=-\frac{4 |S|^4 \lsthree^2\lsfour^2}{M^4}+\left(\frac{|S|^2(\lsthree^2+\lsfour^2)}{M^2}+\lxone^2\right)^2\, .
\end{align}
At small $S$ we find
\begin{align}\label{eq:M1SmallSExp}
V^{(1)}_{F^2}(|S| \ll M)&= \frac{F^2 \lxone^2}{16\pi^2} \left[1+ \ln \left(\lxone^2 \frac{M^2}{\Lambda^2}\right)\right]+\frac{F^2}{32\pi^2 \lxone^2} \frac{|S|^4}{M^4} \Bigg[\lsthree^4+8\lsthree^2\lsfour^2+\lsfour^4\notag\\
&\specialcell{\hfill +2\lsthree^2\lsfour^2 \ln\left(\frac{\lsthree^2\lsfour^2}{\lxone^4} \frac{|S|^4}{M^4} \right)\Bigg]+\mc{O}\left( \frac{|S|^6}{M^6} \right)\,.\quad\quad}
\end{align}
Note that there is no term proportional to $|S|^2$.
The stability of the origin is dictated by the sign of the fourth derivative
\begin{align}
\frac{\partial^4 V^{(1)}_{F^2}(|S| \ll M)}{\partial S^2 \partial S^{*2}}&=\frac{F^2}{8\pi^2 \lxone^2 M^4}\left[ \lsthree^4+20 \lsthree^2\lsfour^2+\lsfour^4+2\lsthree^2\lsfour^2 \ln\left(\frac{\lsthree^2\lsfour^2}{\lxone^4} \frac{|S|^4}{M^4} \right)\right]\,.
\end{align}
This is negative for $|S|/M \to 0$, as the logarithmic term dominates over the other positive contributions. Thus we have a local maximum at the origin.

At large $|S|$ we have
\begin{align}
	V^{(1)}_{F^2}\left(|S| \gg M \right)&=\frac{F^2\lxone^2}{16\pi^2} \ln\left(\frac{|S|^2}{\Lambda^2}\right)+\mc{O}\left(\frac{M^2}{|S|^2}\right)\,,
\end{align}
so the potential increases monotonically. Therefore this model has a minimum away from the origin. From \cref{eq:M1SmallSExp}, setting $\frac{\partial V^{(1)}}{\partial |S|} = 0$, we find
 \begin{align}\label{eq:M1VEVsmallS}
	\langle |S| \rangle& \simeq \pm M \frac{\lxone}{\left(\lsthree \lsfour\right)^{1/2}} \exp\left[-\frac{1}{4}\left(5+\frac{\lsthree^4+\lsfour^4}{2\lsthree^2\lsfour^2}\right)\right].
\end{align}
The equality is not exact as we used the expansion for $|S| \ll M$, but this result is useful for some analytic guidance. 
Setting $\lsthree=\lsfour=\lambda_s$, $\lxone= \lambda_x$ and defining
\be \label{eq:xdef}
x \equiv \frac{\lambda_x}{\lambda_s} \, ,
\ee 
we have
\begin{align} \label{eq:approxmin}
	\langle |S| \rangle& \simeq \pm x \ e^{-3/2}M.
\end{align}

\begin{figure}[!t]
\centering
  \includegraphics{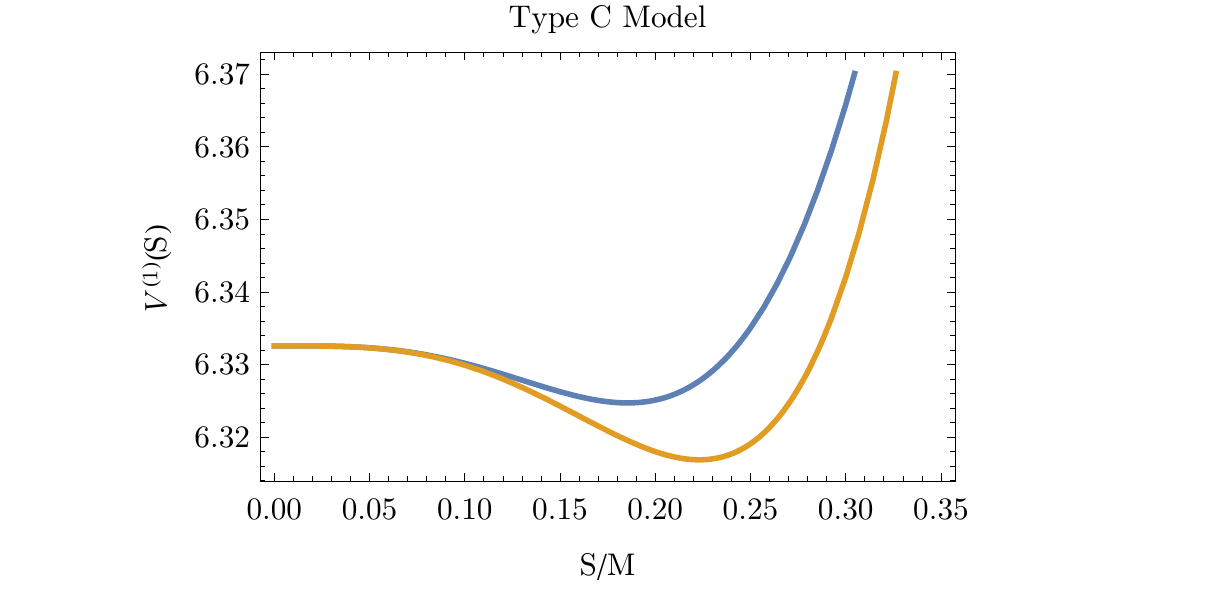}
\caption{A plot of the full K\"ahler effective potential (blue, \cref{eq:full1}) and its expansion around the origin (brown, \cref{eq:M1SmallSExp}) for the model of type C. We set $\lxone=\lsthree=\lsfour=1$, $\Lambda=M$ and $F = 10^{-3} M^2$. }
\label{fig:M1SerVsFull}
\end{figure}

We plot in \cref{fig:M1SerVsFull} the full one-loop potential in blue and its expansion around the origin, \cref{eq:M1SmallSExp}, in brown. We see that the actual minimum is at a slightly lower value than that of \cref{eq:M1VEVsmallS}. Its location can be adjusted by playing with the couplings and using the intuition gained from \cref{eq:approxmin}. 

This model has the desired behavior, as it generates a PQ scale, $f_a = \langle S \rangle$, of the order of the messenger scale.  

\subsubsection{Type D Model}
\begin{figure}
\centering
  \includegraphics{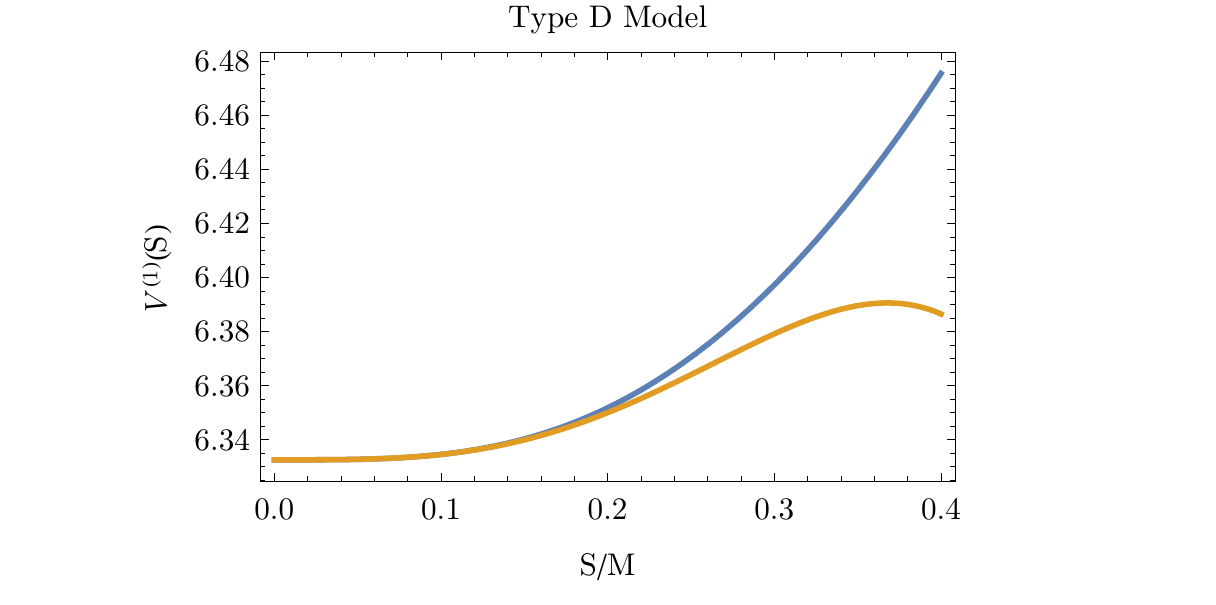}
\caption{A plot of the full K\"ahler effective potential (blue, \cref{eq:full3}) and the expansion of this potential around the origin (brown, \cref{eq:M3SmallSExp}) for model III. The choice of parameters for this plot correspond to $\lxone=\lsthree=\lsfour=1$, $M=\Lambda=1$ and $F = 10^{-3} M^2$. Note that here, for $S/M > 0.2$ the expansion ceases to be a good approximation to the full expression.}
\label{fig:M3SerVsFull}
\end{figure}

From \cref{eq:WPQ3} we have
\be
\mathcal{M}^{D} = 
\begin{pmatrix}
\lxone X & 0 \\
\lsfour S & \lstwo S
\end{pmatrix}\, .
\ee
This model is equivalent to the one considered in Ref.~\cite{Jeong:2011xu}.
The one-loop potential at order $F^2$ is
\begin{align}
	V^{(1)}_{F^2} &= \frac{\lxone^2 F^2}{32\pi^2}\Bigg[\frac{2C}{B}+\left(1-\frac{D}{B^{3/2}}\right)\ln\left(\frac{1}{2\Lambda^2}\left(A-B^{1/2}\right)\right)\notag\\
	&+\left(1+\frac{D}{B^{3/2}}\right)\ln\left(\frac{1}{2\Lambda^2}\left(A+B^{1/2}\right)\right) \Bigg],
	\label{eq:full3}
\end{align}
where
\begin{align}
	A&=|S|^2(\lsfour^2+\lstwo^2)+M^2 \lxone^2,\\
	B&=|S|^4(\lsfour^2+\lstwo^2)^2+2\lxone^2M^2|S|^2(\lsfour^2+\lstwo^2)+M^4 \lxone^4,\\
	C&=\lstwo^2 |S|^4(\lsfour^2+\lstwo^2)+M^2|S|^2(\lsfour^2-2\lstwo^2)\lxone^2+M^4\lxone^4,\\
	D&=|S|^6(\lsfour^2 - \lstwo^2)(\lsfour^2 + \lstwo^2)^2-\lxone^2M^2|S|^4(3\lsfour^4+2\lsfour^2\lstwo^2+3\lstwo^4)\notag\\
	&+3\lxone^4M^4|S|^2(\lsfour^2-\lstwo^2)+\lxone^6M^6.
\end{align}
At small $|S|$             
\begin{align}
V^{(1)}_{F^2}(S\ll M)&= \frac{\lxone^2 F^2}{16\pi^2} \left[ 1+ \ln\left(\lxone^2 \frac{M^2}{\Lambda^2}\right)\right]\notag\\
&\pushright{+\frac{\lsfour^2 F^2 |S|^4}{32\pi^2 \lxone^2 M^4} \left[ \lsfour^2-4\lstwo^2-\lstwo^2\ln\left( \frac{\lstwo^4}{\lxone^4}\frac{|S|^4}{M^4}\right)\right]+\mc{O}\left(\frac{|S|^6}{M^6}\right)\,.}
\label{eq:M3SmallSExp}
\end{align}
The first non zero derivative is
\begin{align}
\frac{\partial^4 V^{(1)}_{F^2}(|S| \ll M)}{\partial S^2 \partial S^{*2}}&=\frac{F^2 \lsfour^2}{8\pi^2 \lxone^2 M^4}\left[ \lsfour^2 -10 \lstwo^2 - \lstwo^2 \ln\left(\frac{\lstwo^4 |S|^4}{\lxone^4 M^4}  \right)\right]\,.
\end{align}
This is positive for $|S|/M \to 0$, where the logarithmic term dominates. Thus we have a local minimum at the origin.
At large $|S|$ we have
\begin{align}
	V^{(1)}_{F^2}\left(|S| \gg M \right)=\frac{F^2\lxone^2}{16\pi^2 (\lsfour^2 + \lstwo^2)} & \left[ 2 \lstwo^2 + \lsfour^2 \ln \left(\frac{|S|^2 (\lsfour^2 + \lstwo^2)}{\Lambda^2} \right)  \right. \notag\\
&	\left.  +\lstwo^2 \ln \left(\frac{M^2 \lstwo^2 \lxone^2 }{e \Lambda^2 (\lsfour^2 + \lstwo^2)} \right)   \right]  +\mc{O}\left(\frac{M^2}{|S|^2}\right)\,,
\end{align}
so the potential increases monotonically. We then infer that the minimum at the origin should be the global one. This can also be seen graphically in \cref{fig:M3SerVsFull}.

We conclude, based on our analysis at one loop, that this model does not generate a large VEV for the field $S$. The authors of Ref.~\cite{Jeong:2011xu} studied the same model at two loops and concluded instead that a large $S$-VEV could be generated.

\section{Anomalies and discrete symmetries}\label{sec:anom}
In this Appendix we discuss the anomalies related to the DGS $\mathbb{Z}_N^{\rm PQ} \times \mathbb{Z}_M^R$ introduced in \cref{sec:quality} to address the issue of the axion quality.\footnote{Anomalies in the discrete symmetries used to protect the axion potential are considered in generality in non supersymmetric models in Ref.~\cite{Cheung:2010hk}.}

As we stated below \cref{eq:CPconstraint} the orders $M$ and $N$ of the DGS must be chosen such that the expressions
\begin{subequations}\label{eq:DGSeqs}
\begin{alignat}{2}
\mathbb{Z}_N^{\rm PQ} &: \qquad \qquad \qquad \qquad \qquad \quad  b\,q_S &&=0 \Mod{N} \,, \\
\mathbb{Z}_M^R &: \qquad \:\: a\, q_X^R + b\, q_S^R = 2a + b\, q_S^R &&=0 \Mod{M}\,, 
\end{alignat}
\end{subequations}
are only satisfied for combinations of $a$ and $b$ such that $a+b\geq8$. In the above expressions $q_\Phi$ and $q^R_\Phi$ label the charges of the chiral superfield $\Phi$ under the $\mathbb{Z}_N^{\rm PQ}$ and $\mathbb{Z}_M^R$ symmetries respectively. Without loss of generality we can choose the charge of the superspace coordinate $q^R_\theta =1$; it follows that $q^R_X = 2$. For the sake of simplicity we take $X$ uncharged under the $\mathbb{Z}_N^{\rm PQ}$ symmetry, $q_X = 0$. From \cref{eq:DGSeqs} a minimal choice of $N$ and $M$, such that the constraint $a+b\geq8$ is satisfied, would be $N=3$ and $M=5$, with $\rcharge{S}=1$.

Let us now consider the model defined by \cref{eq:fullmod}. The form of the superpotential constrains the charges as follows
\begin{subequations}
\begin{align}
\dcharge{X}+ \dcharge{\Phi_1} + \dcharge{\bar{\Phi}_1} &= 0 \Mod{N}\,,\\
\dcharge{S}+ \dcharge{\Phi_1}+\dcharge{\bar{\Phi}_2} &= 0 \Mod{N} \,, \label{eq:dis1}\\
\dcharge{S}+ \dcharge{\Phi_2}+\dcharge{\bar{\Phi}_1} &= 0 \Mod{N}\, , \label{eq:dis2} \\
\rcharge{X}+ \rcharge{\Phi_1} + \rcharge{\bar{\Phi}_1} &= 2\rcharge{\theta} \Mod{M}\,,\\
\rcharge{S}+ \rcharge{\Phi_1}+\rcharge{\bar{\Phi}_2} &= 2\rcharge{\theta} \Mod{M} \,, \label{eq:disR1} \\
\rcharge{S}+ \rcharge{\Phi_2}+\rcharge{\bar{\Phi}_1} &= 2\rcharge{\theta} \Mod{M}\, \label{eq:disR2} .
\end{align}
\end{subequations}
For the anomalies we are interested in the charges of the fermions, which we denote with the subscript $f$ next to the field subscript. Summing \cref{eq:dis1} and \cref{eq:dis2} we get
\begin{align} \label{eq:PQdis}
\sum_i \dcharge{\Phi_i,f}= \sum_i \dcharge{\Phi_i} = -2\dcharge{S} \Mod{N}\, ,
\end{align}
while summing \cref{eq:disR1} and \cref{eq:disR2} we get
\begin{align} \label{eq:Rdis}
\sum_i \rcharge{\Phi_i,f} = \sum_i \left(\rcharge{\Phi_i}-\rcharge{\theta}\right) = 2(2\rcharge{\theta}-\rcharge{S}-2\rcharge{\theta})=-2\rcharge{S}\Mod{M} \,.
\end{align}
In these expressions the index $i$ runs over all the messenger fields in the model.
The requirement that our discrete symmetries be anomaly free amounts to \footnote{We follow here a notation similar to that of Appendix B in Ref.~\cite{Lee:2011dya}. In principle one should also worry about the anomaly coefficients $A_{(\mathbb{Z}_N)^3}$ and $A_{\text{grav}-\text{grav}-\mathbb{Z}_N}$. However, as was argued in Ref.~\cite{Banks:1991xj}, this quantity is particularly sensitive to heavy fractionally charged high energy states. For this reason we do not consider it here. Such arguments also hold when the $\mathbb{Z}_N$ symmetry is promoted to a discrete $R$-symmetry.}
\begin{subequations}
\begin{align}
 A_{G-G-\mathbb{Z}_N^{\rm PQ}}&=\sum_{\boldsymbol{r}_{f}} \ell\left(\boldsymbol{r}_{f}\right) \, \dcharge{\Phi_i,f} = 0 \Mod{\eta_N} \;, \\
 A_{G-G-\mathbb{Z}^R_M} &= \sum_{\boldsymbol{r}_{f}} \ell\left(\boldsymbol{r}_{f}\right) \, \rcharge{\Phi_i,f} + \ell\left(\text{adj}\,G\right) \, \rcharge{\theta} = 0 \Mod{\eta_M} \;,
\end{align}
\end{subequations}
where 
\begin{align}
 \eta_N&=\left\{\begin{array}{ll}
    N & \text{for $N$ odd}\;,\\
    N/2 & \text{for $N$ even}\;,
 \end{array}\right.
\end{align}
The first sum runs over all irreducible representations $\boldsymbol{r}_f$ of the gauge group $G$ with Dynkin index
$\ell(\boldsymbol{r}_f)$, while the second sum runs over all fermions. In our conventions $\ell(\boldsymbol{N})=1/2$ for the fundamental representation of \SU{N}, and $\ell\left(\text{adj}\,G\right) = N$ for the adjoint representation.

For the PQ symmetry, using \cref{eq:PQdis} we obtain
\begin{align}
 A_{\text{SU}(5)-\text{SU}(5)-\mathbb{Z}_N^{\rm PQ}}&=-\dcharge{S}= 0 \Mod{\eta_N} \; .
\end{align}
As we require $q_S \neq 0$, this is only satisfied for $N$ even and $\dcharge{S}=N/2$. Thus we can choose a $\mathbb{Z}_2^{\rm PQ}$ symmetry which is anomaly free. Then, to ensure a sufficient quality of the PQ symmetry without considering anomaly constraints on $\mathbb{Z}_M^R$, the minimal choice is $\mathbb{Z}_2^{\rm PQ}\times \mathbb{Z}_9^R$, with 
$\rcharge{S}=2$.

For the $R$-symmetry, using \cref{eq:Rdis} we obtain
\begin{align}
 A_{\text{SU}(5)-\text{SU}(5)-\mathbb{Z}^R_M} &= 5-\rcharge{S}= 0 \Mod{\eta_M}  \;, 
\end{align}
where we have included the contribution arising from the $\text{SU}(5)$ gauginos. The only discrete $R$-symmetry for $M\leq 12$ where this anomaly coefficient vanishes and all PQ violating non-renormalizable operators with $a+b < 8$ are forbidden is a $\mathbb{Z}_{11}^R$ symmetry where $\rcharge{S}=5$. These results, as stated in the main text, are dependent upon cancellations in only the $R$ and PQ-breaking sectors resepectively. Namely, they will be different if one includes contributions from the visible sector or additional matter in the hidden sector.



\bibliographystyle{JHEP}
\bibliography{Saxion}

\end{document}